\documentstyle[12pt,epsfig]{article}
\begin{document}
\pagestyle{plain}

%
\newcommand{\be}{\begin{equation}}
\newcommand{\ee}{\end{equation}\noindent}
\newcommand{\bear}{\begin{eqnarray}}
\newcommand{\ear}{\end{eqnarray}\noindent}
\newcommand{\no}{\noindent}
\date{}
\renewcommand{\theequation}{\arabic{section}.\arabic{equation}}
\renewcommand{\arraystretch}{2.5}
\newcommand{\GeV}{\mbox{GeV}}
\newcommand{\cL}{\cal L}
\newcommand{\D}{\cal D}
\newcommand{\Dhalf}{{D\over 2}}
\newcommand{\Det}{{\rm Det}}
\newcommand{\PP}{\cal P}
\newcommand{\G}{{\cal G}}
\def\GBd12{{\dot G}_{B12}}
\def\R{1\!\!{\rm R}}
\def\Eins{\mathord{1\hskip -1.5pt
\vrule width .5pt height 7.75pt depth -.2pt \hskip -1.2pt
\vrule width 2.5pt height .3pt depth -.05pt \hskip 1.5pt}}
\newcommand{\symb}{\mbox{symb}}
\renewcommand{\arraystretch}{2.5}
\newcommand{\slD}{\raise.15ex\hbox{$/$}\kern-.57em\hbox{$D$}}
\newcommand{\slpartial}{\raise.15ex\hbox{$/$}\kern-.57em\hbox{$\partial$}}
\newcommand{\slG}{{{\dot G}\!\!\!\! \raise.15ex\hbox {/}}}
\newcommand{\Gd}{{\dot G}}
\newcommand{\Gund}{{\underline{\dot G}}}
\def\np{n_{+}}
\def\nm{n_{-}}
\def\Np{N_{+}}
\def\Nm{N_{-}}
\def\PITD{{(4\pi T)}^{-{D\over 2}}}
\def\non{\nonumber}
\def\beqn*{\begin{eqnarray*}}
\def\eqn*{\end{eqnarray*}}
\def\sy{\scriptscriptstyle}
\def\footstrut{\baselineskip 12pt}
\def\square{\kern1pt\vbox{\hrule height 1.2pt\hbox{\vrule width 1.2pt
   \hskip 3pt\vbox{\vskip 6pt}\hskip 3pt\vrule width 0.6pt}
   \hrule height 0.6pt}\kern1pt}
\def\slash#1{#1\!\!\!\raise.15ex\hbox {/}}
\def\dint#1{\int\!\!\!\!\!\int\limits_{\!\!#1}}
\def\bra#1{\langle #1 |}
\def\ket#1{| #1 \rangle}
\def\vev#1{\langle #1 \rangle}
\def\rightvac{\mid 0\rangle}
\def\leftvac{\langle 0\mid}
\def\dps{\displaystyle}
\def\sy{\scriptscriptstyle}
\def\half{{1\over 2}}
\def\third{{1\over3}}
\def\fourth{{1\over4}}
\def\fifth{{1\over5}}
\def\sixth{{1\over6}}
\def\seventh{{1\over7}}
\def\eigth{{1\over8}}
\def\ninth{{1\over9}}
\def\tenth{{1\over10}}
\def\pa{\partial}
\def\ddtau{{d\over d\tau}}
\def\ge{\hbox{\textfont1=\tame $\gamma_1$}}
\def\gz{\hbox{\textfont1=\tame $\gamma_2$}}
\def\gd{\hbox{\textfont1=\tame $\gamma_3$}}
\def\go{\hbox{\textfont1=\tamt $\gamma_1$}}
\def\gt{\hbox{\textfont1=\tamt $\gamma_2$}}
\def\gth{\hbox{\textfont1=\tamt $\gamma_3$}} 
\def\gf{\hbox{$\gamma_5\;$}}
\def\ie{\hbox{$\textstyle{\int_1}$}}
\def\iz{\hbox{$\textstyle{\int_2}$}}
\def\id{\hbox{$\textstyle{\int_3}$}}
\def\ldop{\hbox{$\lbrace\mskip -4.5mu\mid$}}
\def\rdop{\hbox{$\mid\mskip -4.3mu\rbrace$}}
\def\eps{\epsilon}
\def\epshalf{{\epsilon\over 2}}
\def\e{\mbox{e}}
\def\g{\mbox{g}}
\def\pa{\partial}
\def\kinb{{1\over 4}\dot x^2}
\def\kinf{{1\over 2}\psi\dot\psi}
\def\expk{{\rm exp}\biggl[\,\sum_{i<j=1}^4 G_{Bij}k_i\cdot k_j\biggr]}
\def\expp{{\rm exp}\biggl[\,\sum_{i<j=1}^4 G_{Bij}p_i\cdot p_j\biggr]}
\def\expshort{{\e}^{\half G_{Bij}k_i\cdot k_j}}
\def\expabb{{\e}^{(\cdot )}}
\def\epseps#1#2{\varepsilon_{#1}\cdot \varepsilon_{#2}}
\def\epsk#1#2{\varepsilon_{#1}\cdot k_{#2}}
\def\kk#1#2{k_{#1}\cdot k_{#2}}
\def\G#1#2{G_{B#1#2}}
\def\Gp#1#2{{\dot G_{B#1#2}}}
\def\GF#1#2{G_{F#1#2}}
\def\Dab{{(x_a-x_b)}}
\def\Dsq{{({(x_a-x_b)}^2)}}
\def\lag{( -\partial^2 + V)}
\def\4piTD{{(4\pi T)}^{-{D\over 2}}}
\def\4piT4{{(4\pi T)}^{-2}}
\def\TintmD{{\dps\int_{0}^{\infty}}{dT\over T}\,e^{-m^2T}
    {(4\pi T)}^{-{D\over 2}}}
\def\Tintm4{{\dps\int_{0}^{\infty}}{dT\over T}\,e^{-m^2T}
    {(4\pi T)}^{-2}}
\def\Tintm{{\dps\int_{0}^{\infty}}{dT\over T}\,e^{-m^2T}}
\def\Tint{{\dps\int_{0}^{\infty}}{dT\over T}}
\def\pint{{\dps\int}{dp_i\over {(2\pi)}^d}}
\def\Dx{\dps\int{\cal D}x}
\def\Dy{\dps\int{\cal D}y}
\def\Dpsi{\dps\int{\cal D}\psi}
\def\Tr{{\rm Tr}\,}
\def\tr{{\rm tr}\,}
\def\sumij{\sum_{i<j}}
\def\freeexp{{\rm e}^{-\int_0^Td\tau {1\over 4}\dot x^2}}
\def\arraystretch{2.5}
\def\Ge{\mbox{GeV}}
\def\dA{\partial^2}
\def\DA{\sqsubset\!\!\!\!\sqsupset}
\def\FFdual{F\cdot\tilde F}
%
%
\def\bbbr{{\rm I\!R}}
\def\bbbone{{\mathchoice {\rm 1\mskip-4mu l} {\rm 1\mskip-4mu l}
{\rm 1\mskip-4.5mu l} {\rm 1\mskip-5mu l}}}
\def\bbbz{{\mathchoice {\hbox{$\sf\textstyle Z\kern-0.4em Z$}}
{\hbox{$\sf\textstyle Z\kern-0.4em Z$}}
{\hbox{$\sf\scriptstyle Z\kern-0.3em Z$}}
{\hbox{$\sf\scriptscriptstyle Z\kern-0.2em Z$}}}}
\pagestyle{empty}
\renewcommand{\thefootnote}{\fnsymbol{footnote}}
\hskip 9cm {\sl LAPTH-758/99}
\vskip-.1pt
\vskip .4cm
\begin{center}
{\Large\bf Vacuum Polarisation Tensors in Constant 
Electromagnetic Fields: Part I}
\vskip1.3cm

\vskip.5cm
 {\large Christian Schubert
}
\\[1.5ex]
{\it
Laboratoire d'Annecy-le-Vieux
de Physique Th{\'e}orique LAPTH\\
Chemin de Bellevue,
BP 110\\
F-74941 Annecy-le-Vieux CEDEX\\
FRANCE\\
schubert@lapp.in2p3.fr\\
}
\vskip.5cm

\vskip 2.5cm
 {\large \bf Abstract}
\end{center}
\begin{quotation}
\noindent
The string-inspired technique is used for
the calculation of vacuum polarisation tensors in
constant electromagnetic fields.
In the first part of this series, we
give a detailed exposition of the
method for the case of the QED one-loop
N-photon amplitude  
in a general constant electromagnetic
background field.
The two-point cases are calculated explicitly,
leading to compact representations for
the constant field vacuum polarisation tensors
for both scalar and spinor QED.

\end{quotation}
\clearpage
\renewcommand{\thefootnote}{\protect\arabic{footnote}}
\pagestyle{plain}

\setcounter{page}{1}
\setcounter{footnote}{0}

\section{Introduction: QED 
Processes in Constant Electromagnetic Fields}
\renewcommand{\theequation}{1.\arabic{equation}}
\setcounter{equation}{0}

Processes involving constant electromagnetic fields play a special
role in quantum electrodynamics. An obvious physical reason is
that in many cases a general field can be treated as a constant
one to a good approximation. In QED this is expected to be the
case if the variation of the field is small on the scale of the
electron Compton wavelength.
Mathematically, the constant
field is distinguished by being 
one of the very few known 
field configurations for which the Dirac equation
can be solved exactly, allowing one to obtain results which are
nonperturbative in the field strength. An early and
well-known example is the Euler -- Heisenberg Lagrangian
\cite{eulkoc,eulhei,weisskopf}, 
the one-loop QED vacuum amplitude in a
constant field, 

\bear
{\cal L} &=& {1\over 8\pi^2}\int_0^{\infty}{ds\over s}
\,\e^{-ism^2}\,
e^2ab{\cos(eas)\cosh(ebs)\over \sin(eas)\sinh(ebs)}
\label{eulhei}
\ear\no
where $a^2-b^2 \equiv {\bf B}^2 - {\bf E}^2$,
$ab \equiv {\bf E}\cdot{\bf B}$.
This Lagrangian encodes the information on the
low energy limit of the one-loop photon S - matrix
in a form which is convenient for the derivation
of nonlinear QED effects such as photon -- photon
scattering and vacuum birefringence
\cite{ritusbook,ditreu,grereibook}.
Vacuum birefringence is a subject of actual interest
\cite{ditgie}
since, due to recent improvements in laser technology,
the first measurement of this effect
in the laboratory seems now imminent \cite{pvlas}.

Schwinger's
ingenious use of Fock's proper--time
method in 1951
~\cite{fock,schwinger51} 
allowed him to reproduce this
result, as well as the analogous one for
scalar QED, with considerably less effort. 
Shortly later Toll \cite{toll} initiated the 
study of the effect of 
a background field on the one-loop photon propagator.
This subject was then over the years investigated by
a number of authors, first for the pure magnetic field 
\cite{minguzzi,baibre,constantinescu,tsaerb,covkal,shabad}
and crossed field cases \cite{narozhnyi,ritus72}.
The vacuum polarisation tensor in a general electromagnetic
field was first obtained in \cite{batsha} and
given in more explicit form in \cite{bakast,urrutia,artimovich}.
The recent \cite{gies} contains another recalculation of this
quantity, as well as a detailed analysis of the implications
for light propagation.

Another consequence of the 
presence of a background field is the invalidation of 
Furry's theorem;
already the three-photon amplitude is non-vanishing in
a constant field.
Moreover the modification of the photon dispersation relation
through the background field can, depending on the photon
polarisations, lead to the opening up of phase space for
the photon splitting process $\gamma \to \gamma + \gamma$.
For the case of a magnetic field this process was calculated in the
low photon energy limit in \cite{biabia} and for 
general photon energies in \cite{adler71}. The amplitude
turns out to be very small for the magnetic field strengths
presently attainable in the laboratory.
Nevertheless, the photon
splitting process is believed to be of relevance
for the physics of neutron stars which are known
to have magnetic fields approaching, and even surpassing,
the ``critical'' magnetic field strength
$B_{\rm crit} = {m_e^2\over e} =
4.41\times 10^{13}$ Gauss
\cite{raffelt,barhar,kaspi}.
It seems also not impossible that, with some further
improvements in laser technology, photon splitting may
be observable in the laboratory in the near future
\cite{bakalov,fermilab877}.

For QED calculations in constant external fields it is possible 
and advantageous to take account of the 
field already at the level of the Feynman rules,
i.e. to absorb it into the free electron propagator.
Suitable formalisms have been developed
decades ago ~\cite{geheniau,tsai,bakast}.
However beyond the simplest special cases
they lead to exceedingly tedious and
cumbersome calculations.

A different and more efficient
formalism for such calculations 
has been developed during the last few
years, using the so-called ``string-inspired'' technique.
The idea of using string theory methods as a 
practical tool
for calculations in ordinary quantum field theory
was 
advocated by Bern and Kosower \cite{berkos}.
Their work led to the formulation of new computation
rules for QCD amplitudes which made it,
for example, feasible
to perform a complete calculation of the
one-loop five gluon amplitudes \cite{5glu}.
A parallel line of work led to the formulation
of analogous computation rules for quantum gravity
\cite{bedush,dunnor}.
The relation between the string-derived rules and ordinary
Feynman rules was clarified in \cite{berdun}.
 
Later it was found that even in abelian gauge theory
significant improvements over standard field theory
methods can be obtained along these lines 
\cite{strassler,ss3}. 
Moreover, in this formalism the inclusion of constant external fields 
turned out to require only relatively minor modifications
\cite{ss1,cadhdu,gussho,shaisultanov,rss}.
For this reason it has been extensively applied
to constant field processes in QED.
This includes a recalculation of the photon-splitting amplitude
\cite{adlsch} as well as high order calculations of the
derivative expansion of the QED effective action 
\cite{cadhdu,gussho,shovkovy,fhss}.
A generalization to multiloop photonic
amplitudes \cite{ss2,ss3,dashsu,rolsat,rss} 
was applied to a calculation
of the two-loop correction to the Euler-Heisenberg
Lagrangian, using both proper-time \cite{rss}
and dimensional regularisation \cite{frss,korsch,dunsch}.
We will not discuss here the involved history
of this subject but rather refer the interested reader to the
review articles \cite{berntasi,zako}. For other
work on QED amplitudes similar
in spirit to the string-inspired approach see
\cite{fried,baboca,bafrsh,rajeev,frgish,mckshe}.

In the present paper we first give, in chapter 2,
a detailed and self-contained 
exposition of the string-inspired technique for the
calculation of one-loop scalar/spinor 
QED photon amplitudes in
vacuum. 
In chapter 3 we extend this formalism to the inclusion
of constant external fields along the lines of
\cite{rss}. 
As a technical improvement on \cite{rss}
we derive a decomposition of the generalised worldline
Green's functions ${\cal G}_{B,F}$ in terms of
the matrices $\Eins,F,\tilde F, F^2$ 
with coefficients that are functions of the two
standard Maxwell invariants.
This will allow us to
arrive at explicit results in a manifestly Lorentz
covariant way. In chapter 4 we apply the formalism
to a calculation of the scalar and spinor QED vacuum
polarisation tensors in a constant field. 
Chapter 5 contains our conclusions.

\section{The QED One-Loop $N$-Photon Amplitude in Vacuum} 
\renewcommand{\theequation}{2.\arabic{equation}}
\setcounter{equation}{0}

In \cite{strassler} it was shown that, for the QED case,
the full content of the Bern-Kosower rules can
be captured using an approach to quantum field theory
based on first-quantized particle path integrals
(`worldline path integrals').

\subsection{Scalar Quantum Electrodynamics}

For the case of scalar QED the basic formulas
needed go back to Feynman \cite{feyn}.
The one-loop effective action due to a scalar loop for a Maxwell
background can (in modern notation) be written as
\footnote{%
We work initially in the Euclidean with
a positive definite metric
$g_{\mu\nu}={\,\mathrm diag}(++++)$.
The Euclidean field strength tensor is defined by
$F^{ij}= \varepsilon_{ijk}B_k, i,j = 1,2,3$,
$F^{4i}=-iE_i$, its dual by
$\tilde F^{\mu\nu} = \half 
\varepsilon^{\mu\nu\alpha\beta}F^{\alpha\beta}$
with $\varepsilon^{1234} = 1$.  
The corresponding Minkowski space amplitudes
can be obtained by replacing 
$g_{\mu\nu}\rightarrow \eta_{\mu\nu}
= {\,\mathrm diag}(-+++)$, $
k^4\rightarrow -ik^0, T\rightarrow is,
\varepsilon^{1234}
\rightarrow
i\varepsilon^{1230},
\varepsilon^{0123}=1,
F^{4i}\rightarrow F^{0i}=E_i,
\tilde F^{\mu\nu}\rightarrow -i\tilde F^{\mu\nu}$.
}

\bear
\Gamma_{\rm scal}[A]
&=&
\Tintm
\int_{x(T)=x(0)}{\cal D}x(\tau)
\,\e^{-\int_0^T d\tau\Bigl(
\kinb
+ie\,\dot x\cdot A(x(\tau))
\Bigr)}
\non\\
\label{scalarqedpi}
\ear\no
Here $T$ denotes the usual proper-time for the
loop fermion. For fixed $T$
$\int {\cal D}x$ denotes an integral over the space of all closed
loops in spacetime with periodicity $T$.

This path integral can be used for the calculation of the
effective action itself as well as for obtaining
the corresponding scattering amplitudes.
As a first step in any evaluation,
one has to take care of the zero mode
contained in it. This is done by fixing the
average position of the loop, i.e. one writes

\bear
x^{\mu}(\tau) &=& x_0^{\mu} + y^{\mu}(\tau)\non\\
\Dx &=& \int dx_0 \Dy\non\\
\label{split}
\ear\no
where

\begin{equation}
x_0^{\mu}\equiv {1\over T}\int_0^T d\tau\, x^{\mu}(\tau)
\label{defx0}
\end{equation}
\no
The remaining $y$ 
path integral is, in the `string-inspired formalism',
performed using the Wick contraction rule

\begin{eqnarray}
\langle y^\mu(\tau_1)\,y^\nu(\tau_2)\rangle
&=& -g^{\mu\nu}{G}_{B}(\tau_1,\tau_2)\label{corry}
\end{eqnarray}\no
where 

\begin{eqnarray}
G_B(\tau_1,\tau_2)
    &=& \mid \tau_1 - \tau_2\mid -
{{(\tau_1 - \tau_2)}^2\over T}\non\\
\label{defGB}
\end{eqnarray}
\no
A ``dot'' always refers to a derivative in the
first variable. 

The free Gaussian path integral
determinants are, in our conventions, given by

\bear
\int{\cal D}y
\,\e^{-\int_0^T d\tau\, \fourth
\dot y^2}
&=& {(4\pi T)}^{-{D\over 2}} \label{freepiy}
\ear
Here $D$ denotes the spacetime dimension. Although in
this paper we will consider only the four-dimensional
case in this factor $D$ must be left variable in
anticipation of dimensional regularization.

We can use this path integral for constructing the
scalar QED $N$-photon amplitude as follows \cite{strassler}.
Expanding the
`interaction exponential',

\bear
{\rm exp}\Bigl[
-\int_0^Td\tau\, ieA_{\mu}\dot x^{\mu}
\Bigr]
&=&\sum_{N=0}^{\infty}
{{(-ie)}^N\over N!}
\prod_{i=0}^N
\int_0^Td\tau_i
\biggl[
\dot x^{\mu}(\tau_i)
A_{\mu}(x(\tau_i))
\biggr]
\label{expandint}
\non\\
\ear\no
the individual terms correspond to Feynman diagrams
describing a fixed number of
interactions of the scalar loop with
the external field.
The corresponding $N$ -- photon
scattering amplitude is then obtained by
specializing to a background
consisting of 
a sum of plane waves with definite
polarizations,

\be
A_{\mu}(x)=
\sum_{i=1}^N
\varepsilon_{i\mu}
\e^{ik_i\cdot x}
\label{planewavebackground}
\ee\no
and picking out the term containing every
$\varepsilon_i$ once.
This immediately yields the following representation for
the $N$ - photon amplitude, 

\bear
\Gamma_{\rm scal}[k_1,\varepsilon_1;\ldots;k_N,\varepsilon_N]
&=&
(-ie)^{N}
\Tintm 
\PITD
\non\\&&
\hspace{-21pt}
\times
\Bigl\langle
V_{A}^{0}[k_1,\varepsilon_1]\ldots
V_{A}^{0}[k_N,\varepsilon_N]
\Bigr\rangle
\non\\
\label{repNvector}
\ear\no
Here $V_A^{0}$ denotes 
the same photon
vertex operator which is also used in string perturbation
theory (see, e.g., \cite{grscwi}),

\begin{equation}
V_A^{0}[k,\varepsilon]
\equiv
\int_0^Td\tau\,
\varepsilon\cdot \dot x(\tau)
\,{\rm e}^{ikx(\tau)}
\label{photonvertopscal}
\end{equation}
\noindent
At this stage the zero-mode integration
(\ref{split}) can be performed, 
yielding the energy-momentum
conservation factor

\bear
\int d^Dx_0 \prod_{i=1}^N \e^{ik_i\cdot x_0}
&=&
{(2\pi )}^D\delta (\sum k_i)
\label{momcons}
\ear\no
The reduced path integral $\int{\cal D}y$ is Gaussian.
Its evaluation therefore amounts to Wick contracting the 
expression

\be
\biggl\langle
\dot y_1^{\mu_1}\e^{ik_1\cdot y_1}
\cdots
\dot y_N^{\mu_N}\e^{ik_N\cdot y_N}
\biggr\rangle
\label{scalqedwick}
\ee
\no
using the correlator (\ref{corry}).
For the performance of the Wick contractions it is
convenient to formally exponentiate all the $\dot y_i$'s, 
writing

\be
\varepsilon_i\cdot
\dot y_i\,\e^{ik_i\cdot y_i}
=
\e^{\varepsilon_i\cdot\dot y_i
+ik_i\cdot y_i}
\mid_{{\rm lin}(\varepsilon_i)}
\label{formexp}
\ee
\no
This allows one to rewrite the product of $N$ photon vertex
operators as an exponential. Then one needs only
to `complete the square' to arrive at the following
closed expression for the one-loop
$N$ - photon amplitude \cite{strassler}

\begin{eqnarray}
\Gamma_{\rm scal}[k_1,\varepsilon_1;\ldots;k_N,\varepsilon_N]
&=&
{(-ie)}^N
{(2\pi )}^D\delta (\sum k_i)
\non\\
&&\hspace{-120pt}
\times {\dps\int_{0}^{\infty}}
{dT\over T}
{(4\pi T)}^{-{D\over 2}}
e^{-m^2T}
\prod_{i=1}^N \int_0^T 
d\tau_i
\nonumber\\
&&\hspace{-130pt}
\times
\exp\biggl\lbrace\sum_{i,j=1}^N 
\bigl\lbrack \half G_{Bij} k_i\cdot k_j
+i\dot G_{Bij}k_i\cdot\varepsilon_j 
+\half\ddot G_{Bij}\varepsilon_i\cdot\varepsilon_j
\bigr\rbrack\biggr\rbrace
\mid_{\rm multi-linear}.
\nonumber\\
\label{scalarqedmaster}
\end{eqnarray}
\no
Here it is understood that only the terms linear
in all the $\varepsilon_1,\ldots,\varepsilon_N$
have to be taken. 
Besides the Green's function $G_B$ also its first and
second deriatives appear,

\begin{eqnarray}
\dot G_B(\tau_1,\tau_2) &=& {\rm sign}(\tau_1 - \tau_2)
- 2 {{(\tau_1 - \tau_2)}\over T}\nonumber\\
\ddot G_B(\tau_1,\tau_2)
&=& 2 {\delta}(\tau_1 - \tau_2)
- {2\over T}\quad \nonumber\\
\label{GdGdd}
\end{eqnarray}
\noindent
With `dots' we generally denote a
derivative acting on the first variable,
$\dot G_B(\tau_1,\tau_2) \equiv {\partial\over
{\partial\tau_1}}G_B(\tau_1,\tau_2)$, 
and we abbreviate
$G_{Bij}\equiv G_B(\tau_i,\tau_j)$ etc.

The expression (\ref{scalarqedmaster})
is identical with the corresponding special case
of the `Bern-Kosower Master Formula' \cite{berkos}.
Let us consider explicitly the vacuum polarisation case,
$N=2$. For $N=2$ the expansion of the exponential
factor 
yields the following expression,

\be
\Bigl(
\ddot G_{B12}\varepsilon_1\cdot\varepsilon_2
+
\dot G_{B12}^2
\varepsilon_1\cdot k_2
\varepsilon_2\cdot k_1
\Bigr)
{\rm e}^{G_{B12}k_1\cdot k_2}
\non\\
\label{N=2wick}
\ee\no
After performing
a partial integration on the first
term of eq. (\ref{N=2wick}) in either
$\tau_1$ or $\tau_2$, the integrand 
turns into

\be
\Bigl(
\varepsilon_1\cdot\varepsilon_2
k^2
-
\varepsilon_1\cdot k
\varepsilon_2\cdot k
\Bigr)
\dot G_{B12}^2
{\rm e}^{-G_{B12}k^2}
\label{N=2partint2}
\ee\no
($k=k_1=-k_2$).
Thus we have

\bear
\Gamma_{\rm scal}[k_1,\varepsilon_1;k_2,\varepsilon_2]
&=&
{(2\pi )}^D\delta (k_1+k_2)
\varepsilon_1\cdot\Pi_{\rm scal}(k)\cdot\varepsilon_2\non\\
\Pi_{\rm scal}^{\mu\nu}(k)
&=&
{e^2\over {(4\pi )}^{D\over 2}}
(k^{\mu}k^{\nu}-g^{\mu\nu}k^2)
\int_0^{\infty}
{dT\over T}
T^{-{D\over 2}}
e^{-m^2T}
\non\\&&\times
\prod_{i=1}^2 \int_0^T 
d\tau_i
\,\dot G_{B12}^2
{\rm e}^{-G_{B12}k^2}
\label{scalvpprel}
\ear\no
Note that the transversality of the
vacuum polarization tensor
is already manifest.
We rescale to the unit circle, 
$\tau_i = Tu_i, i = 1,2$, and use translation
invariance in $\tau$ to fix the zero to 
be at the location of the second vertex operator,
$u_2=0, u_1=u$.
We have then

\bear
G_B(\tau_1,\tau_2)&=&Tu(1-u),\quad
\dot G_B(\tau_1,\tau_2)=1-2u
\nonumber\\
\label{scaledown}
\ear\no
After performing the trivial $T$ - integration
one arrives at

\bear
\Pi^{\mu\nu}_{\rm scal}
(k) &=&
{e^2\over {(4\pi )}^{D\over 2}}
\Bigl(k^{\mu}k^{\nu}-g^{\mu\nu}k^2\Bigr)
\Gamma\bigl(2-{D\over 2}\bigl)
\non\\&&\hspace{-26pt}
\times
\int_0^1du
(1-2u)^2
{\Bigl[
m^2 + u(1-u)k^2
\Bigr]
}^{{D\over 2}-2}
\non\\
\label{scalarvpresult}
\ear\no
The result of the final integration is, of course,
the same as is found
in the standard field theory calculation
for the sum of the corresponding two Feynman diagrams
(see, e.g., \cite{itzzub}).

\subsection{Spinor Quantum Electrodynamics}

The worldline path integral representation (\ref{scalarqedpi})
can be generalized to the spinor QED case in various different
ways. The formulation most suitable to the `stringy' approach
uses Grassmann variables \cite{fradkin66,casalbuoni,bermar},

\begin{eqnarray}
\Gamma_{\rm spin}\lbrack A\rbrack &  
= &- \half {\displaystyle\int_0^{\infty}}
{dT\over T}
e^{-m^2T}
{\displaystyle\int} 
{\cal D} x
{\displaystyle\int}
{\cal D}\psi\nonumber\\
& \phantom{=}
&\times
{\rm exp}\biggl [- \int_0^T d\tau
\Bigl ({1\over 4}{\dot x}^2 + {1\over
2}\psi\cdot\dot\psi
+ ieA\cdot x - ie
\psi\cdot F \cdot\psi
\Bigr )\biggr ]\non\\
\label{spinorpi}
\end{eqnarray}
\noindent
Thus we have, in addition to the same coordinate
path integral as in (\ref{scalarqedpi}), a Grassmann
path integral $\int {\cal D}\psi$ representing the fermion spin.
The boundary conditions on the Grassmann path integral are
antiperiodic, 
$\psi(T)=-\psi(0)$, so that there is no new zero mode.
The appropriate correlator is

\begin{eqnarray}
\langle\psi^{\mu}(\tau_1)\, \psi^{\nu}(\tau_2)\rangle
&=& 
\half g^{\mu\nu}{G}_{F}(\tau_1,\tau_2)\label{corrpsi}
\end{eqnarray}\no
Our normalization for the free Grassmann path integral is

\bear
\int{\cal D}\psi\, \e^{-\int_0^Td\tau \,\half\psi\cdot\dot\psi}
&=& 4\label{freepipsi}
\ear\no
The photon vertex operator 
(\ref{photonvertopscal}) acquires an additional Grassmann
piece, 

\begin{equation}
V_A^{\half}[k,\varepsilon]
\equiv
\int_0^Td\tau
\Bigl(
\varepsilon\cdot \dot x
+2i
\varepsilon\cdot\psi
k\cdot\psi
\Bigr)\,
{\rm e}^{ikx}
\label{photonvertopspin}
\end{equation}
\noindent
Looking again at the vacuum polarization case, 
we need to Wick-contract two copies of the
above vertex operator.
The calculation
of $\int{\cal D}x$ is identical with the scalar QED calculation.
The additional contribution from $\int{\cal D}\psi$ is 

\be
{(2i)}^2
\Bigl\langle
\psi^{\mu}_1\psi_1\cdot k_1
\psi^{\nu}_2\psi_2\cdot k_2
\Bigr\rangle
=-
G_{F12}^2
\Bigl(g^{\mu\nu}k^2-k^{\mu}k^{\nu}\Bigr)
=-
\Bigl(g^{\mu\nu}k^2-k^{\mu}k^{\nu}\Bigr)
\label{spinwick2point}
\ee\no
Taking the free Grassmann path integral normalization
(\ref{freepipsi}) into account, eq.(\ref{scalvpprel})
for the scalar QED vacuum polarisation tensor
generalises to the spinor QED case as follows,

\bear
\Pi_{\rm spin}^{\mu\nu}(k)
&=&
-2
{e^2\over {(4\pi )}^{D\over 2}}
(k^{\mu}k^{\nu}-g^{\mu\nu}k^2)
\int_0^{\infty}
{dT\over T}
T^{-{D\over 2}}
e^{-m^2T}
\non\\&&\times
\prod_{i=1}^2 \int_0^T 
d\tau_i
\,\bigl(\dot G_{B12}^2 -G_{F12}^2\bigr)
\,{\rm e}^{-G_{B12}k^2}
\label{spinvpprel}
\ear\no
Proceeding as before one obtains

\bear
\Pi^{\mu\nu}_{\rm spin}
(k) &=&
8
{e^2\over {(4\pi )}^{D\over 2}}
\Bigl(k^{\mu}k^{\nu}-g^{\mu\nu}k^2\Bigr)
\Gamma\bigl(2-{D\over 2}\bigl)
\non\\&&\hspace{-26pt}
\times
\int_0^1du
u(1-u)
{\Bigl[
m^2 + u(1-u)k^2
\Bigr]
}^{{D\over 2}-2}
\non\\
\label{spinorvpresult}
\ear\no
Remarkably, the explicit calculation of
the Grassmann path integral can be circumvented,
and replaced by the following simple pattern matching rule 
\cite{berkos}.
Writing out the exponential in eq.(\ref{scalarqedmaster})
one obtains an integrand

\be
\exp\biggl\lbrace 
\cdots
\biggr\rbrace
\mid_{\rm multi-linear}\quad
={(-i)}^N
P_N(\dot G_{Bij},\ddot G_{Bij})
\exp\biggl[\half
\sum_{i,j=1}^N G_{Bij}k_i\cdot k_j
\biggr]
\label{defPN}
\ee\no
with a certain polynomial $P_N$ depending on the various 
$\dot G_{Bij},\ddot G_{Bij}$ and on the kinematic invariants.
Now one
removes all second derivatives $\ddot G_{Bij}$ appearing
in $P_N$
by suitable partial integrations in the variables
$\tau_i$,

\be
P_N(\dot G_{Bij},\ddot G_{Bij})
\e^{\half\sum G_{Bij}k_i\cdot k_j}
\quad
{\stackrel{\sy{\rm part. int.}}{\longrightarrow}}
\quad
Q_N(\dot G_{Bij})
\e^{\half\sum G_{Bij}k_i\cdot k_j}
\label{partint}
\ee\no
This is possible for any $N$ \cite{berkos}. 
The result is an alternative integrand for the
scalar QED amplitude involving only
$G_B$ and $\dot G_B$. 
The integrand for the spinor
loop case can then,
up to the global factor of $-2$,
 be obtained from the one for
the scalar loop simply
by replacing every closed cycle of $\dot G_B$'s
appearing in $Q_N$ by its
``worldline supersymmetrization'',

\vspace{-10pt}
\begin{equation}
\dot G_{Bi_1i_2} 
\dot G_{Bi_2i_3} 
\cdots
\dot G_{Bi_ni_1}
\rightarrow 
\dot G_{Bi_1i_2} 
\dot G_{Bi_2i_3} 
\cdots
\dot G_{Bi_ni_1}
-
G_{Fi_1i_2}
G_{Fi_2i_3}
\cdots
G_{Fi_ni_1}
\nonumber\\
\label{subrule}
\end{equation}
Note that an expression is considered a cycle
already if it can be put into cycle form
using the antisymmetry of $\dot G_B$ (e.g.
$\dot G_{B12}\dot G_{B12}=-\dot G_{B12}\dot G_{B21}$).
The replacement is done simultaneously on all cycles.

For $N>3$ the result of the partial integration procedure
is not unique, however the above replacement rule
is valid for all possible results.
In \cite{mepartint} a certain standardized way 
was found for performing the partial integrations
which leads to a canonical, permutation symmetric
and gauge invariant decomposition of the 
QED $N$ - photon amplitudes.

\section{The QED N-Photon Amplitude in a Constant Field}
\renewcommand{\theequation}{3.\arabic{equation}}
\setcounter{equation}{0}

\subsection{Generalization of the Bern-Kosower Master Formula}

The presence of an additional constant external field,
taken in Fock-Schwinger gauge centered at $x_0$
\cite{ss1},
changes the path integral Lagrangian
in eq.(\ref{spinorpi}) only by a term
quadratic in the fields,
$
\Delta{\cal L} = {1\over 2}\,ie\,y^{\mu} F_{\mu\nu}
\dot y^{\nu} - ie\,\psi^{\mu} F_{\mu\nu}\psi^{\nu}
$.
The field can therefore be absorbed by a change of the
free worldline propagators, replacing 
$G_B,\dot G_B,G_F$ by (\cite{cadhdu,shaisultanov,rss}; see also
\cite{mckshe})

\bear
{{\cal G}_B}(\tau_1,\tau_2) &=&
{T\over 2{\cal Z}^2}
\biggl({{\cal Z}\over{{\rm sin}({\cal Z})}}
\,{\rm e}^{-i{\cal Z}\dot G_{B12}}
+ i{\cal Z}\dot G_{B12} - 1\biggr)
\label{calGB}\\
\dot{\cal G}_B(\tau_1,\tau_2) &=&
{i\over {\cal Z}}
\biggl({{\cal Z}\over{{\rm sin}({\cal Z})}}
\,{\rm e}^{-i{\cal Z}\dot G_{B12}}
- 1\biggr)
\label{dotcalGB}\\
{\cal G}_{F}(\tau_1,\tau_2) &=&
G_{F12} {{\rm e}^{-i{\cal Z}\dot G_{B12}}\over {\rm cos}({\cal Z})}
\label{calGF}
\ear
where we have defined ${\cal Z}\equiv e{F}T$. 
These expressions should be understood as power
series in the field strength matrix $F$.
Note that the generalized Green's functions are still
translationally invariant in $\tau$, and thus
functions of $\tau_1 - \tau_2$. By writing them
as functions of the vacuum worldline
Green's functions $\dot G_B, G_F$
we have left the  $\tau$ - dependence
implicit.
This allows us to avoid 
making a case distinction between
$\tau_1 >\tau_2$ and $\tau_1 < \tau_2$ that
would become necessary otherwise
\cite{shaisultanov}.
Note also the symmetry properties

\bear
{\cal G}_B(\tau_1,\tau_2) = 
{\cal G}_B^{T}(\tau_2,\tau_1),
\quad
\dot{\cal G}_B(\tau_1,\tau_2) = 
-\dot{\cal G}_B^{T}(\tau_2,\tau_1),
\quad
{\cal G}_F(\tau_1,\tau_2) = 
-
{\cal G}_F^{T}(\tau_2,\tau_1)
\non\\
\label{symmcalGBF}
\ear\no
Since ${\cal G}_B, {\cal G}_F$
are, in general, nontrivial Lorentz matrices,
the Wick contraction rules eqs.(\ref{corry}),(\ref{corrpsi})
have to be replaced by

\begin{eqnarray}
\langle y^{\mu}(\tau_1)y^{\nu}(\tau_2)\rangle
&=&
-{\cal G}_B^{\mu\nu}(\tau_1,\tau_2)\label{corryF}\\
\langle\psi^{\mu}(\tau_1)\psi^{\nu}(\tau_2)\rangle
&=&
\frac{1}{2}{\cal G}_F^{\mu\nu}(\tau_1,\tau_2)\label{corrpsiF}
\end{eqnarray}
\noindent
Another slight complication compared to the vacuum
case is that,
in contrast to their vacuum counterparts,
${\cal G}_B, \dot {\cal G}_B$, and ${\cal G}_F$
have
non-vanishing coincidence limits. Those are
$\tau$ - independent:

\begin{eqnarray}
{\cal G}_B(\tau,\tau) &=&
{T\over 2{\cal Z}^2}
\biggl({\cal Z}\cot ({\cal Z})- 1\biggr) \label{coincalGB}\\
\dot {\cal G}_B(\tau,\tau) &=& i{\rm cot}({\cal Z})
-{i\over {\cal Z}}\label{coindotcalGB}\\
{\cal G}_F(\tau,\tau) &=& -i\,{\rm tan}({\cal Z})\label{coincalGF}
\end{eqnarray}
\noindent
They are obtained from 
eqs.(\ref{calGB}),(\ref{dotcalGB}),(\ref{calGF}) 
using the rules that

\bear
\dot G_B(\tau,\tau) &=& 0, \quad \dot G_B^2(\tau,\tau) =1
\label{rulecoin}
\ear\no

This is almost all we need to know for computing one-loop
photon scattering amplitudes, or the corresponding
effective action, in a constant overall background field. 
The only further information required at the one--loop
level is the change in the free path integral determinants
due to the external field. 
This change is \cite{ss1}

\bear
\!\!\!\!
{(4\pi T)} ^{-{D\over 2}}
&\rightarrow&
{(4\pi T)}^{-{D\over 2}}
{\rm det}^{-{1\over 2}}
\biggl[{\sin({\cal Z})\over {{\cal Z}}}
\biggr] \quad\qquad{\rm (Scalar\; QED)}
\label{scaldetext}\\
\!\!\!\!
{(4\pi T)}^{-{D\over 2}}
&\rightarrow&
{(4\pi T)}^{-{D\over 2}}
{\rm det}^{-{1\over 2}}
\biggl[{\tan({\cal Z})\over {{\cal Z} }}
\biggr] \quad\qquad{\rm (Spinor\; QED)}
\label{spindetext}
\ear\no
Since those determinants describe the vacuum amplitude in
a constant field they turn out to be, of course, 
just the
proper-time integrands of the  
Euler-Heisenberg-Schwinger formulas
(see (\ref{eulhei})).

Retracing our above calculation of the $N$ - photon path integral
with the external field included we arrive at the
following generalization of eq.(\ref{scalarqedmaster}),
representing the scalar QED $N$ - photon scattering amplitude 
in a constant field \cite{shaisultanov,rss}:

\begin{eqnarray}
&&\Gamma_{\rm scal}
[k_1,\varepsilon_1;\ldots;k_N,\varepsilon_N]
=
{(-ie)}^N
{(2\pi )}^D\delta (\sum k_i)
\non\\
&&\hspace{20pt}\times
{\dps\int_{0}^{\infty}}{dT\over T}
{[4\pi T]}^{-{D\over 2}}
e^{-m^2T}
{\rm det}^{-{1\over 2}}
\biggl[{{\rm sin}({\cal Z})\over {\cal Z}}\biggr]
\prod_{i=1}^N \int_0^T 
d\tau_i
\non\\
&&\hspace{20pt}\times
\exp\biggl\lbrace\sum_{i,j=1}^N 
\Bigl\lbrack \half k_i\cdot {\cal G}_{Bij}\cdot  k_j
-i\varepsilon_i\cdot\dot{\cal G}_{Bij}\cdot k_j
+\half
\varepsilon_i\cdot\ddot {\cal G}_{Bij}\cdot\varepsilon_j
\Bigr\rbrack\biggr\rbrace
\mid_{\rm multi-linear}\quad
\nonumber\\
\label{scalarqedmasterF}
\end{eqnarray}
\no
From this formula it is obvious that adding a constant matrix
to ${\cal G}_B$ will have no effect due to momentum
conservation. We can use this fact to get rid
of the coincidence limit of ${\cal G}_B$, (\ref{coincalGB}),
namely instead of ${\cal G}_B$ one can work with
the equivalent Green's function $\bar{\cal G}_B$,
defined by

\bear
\bar{\cal G}_B(\tau_1,\tau_2)
\equiv
{\cal G}_B(\tau_1,\tau_2) - {\cal G}_B(\tau,\tau)
\label{defbarcalGB}
\ear\no
No such redefinition is possible
for $\dot{\cal G}_B$ or ${\cal G}_F$.

The transition from scalar to spinor QED is done as in the
vacuum case, again with only some minor modifications.
The spinor QED integrand for a given number of
photon legs $N$ is obtained from the scalar QED
integrand by the following generalization
of the Bern-Kosower algorithm:

\begin{enumerate}

\item
{\it Partial Integration:}
After expanding out the exponential in the
master formula (\ref{scalarqedmasterF}), and 
taking the part linear in all $\varepsilon_1,\ldots,
\varepsilon_N$, remove all second derivatives
$\ddot{\cal G}_B$ appearing in the result by
suitable partial integrations in $\tau_1,\ldots,\tau_N$.

\item
{\it Replacement Rule:}
Apply to the resulting new integrand the
replacement rule (\ref{subrule}) with
$\dot G_B, G_F$ substituted by $\dot {\cal G}_B,
{\cal G}_F$.
Since the Green's functions ${\cal G}_B,{\cal G}_F$
are, in contrast to their vacuum counterparts,
non-trivial matrices in the Lorentz indices,
it must be
mentioned here that the 
cycle property is defined solely in terms of the
$\tau$ -- indices, irrespectively of what happens
to the Lorentz indices. For example, the expression

$$
\varepsilon_1\cdot\dot{\cal G}_{B12}\cdot k_2\,
\varepsilon_2\cdot\dot{\cal G}_{B23}\cdot\varepsilon_3\,
k_3\cdot\dot{\cal G}_{B31}\cdot k_1\,
$$

\noindent
would have to be replaced by

$$
\varepsilon_1\cdot\dot{\cal G}_{B12}\cdot k_2\,
\varepsilon_2\cdot\dot{\cal G}_{B23}\cdot\varepsilon_3\,
k_3\cdot\dot{\cal G}_{B31}\cdot k_1\,
-
\varepsilon_1\cdot {\cal G}_{F12}\cdot k_2\,
\varepsilon_2\cdot {\cal G}_{F23}\cdot\varepsilon_3\,
k_3\cdot {\cal G}_{F31}\cdot k_1
$$

\noindent
The only other difference
compared to the vacuum case is due to the 
non-vanishing coincidence limits
of $\dot{\cal G}_B,{\cal G}_F$,
eqs.(\ref{coindotcalGB}),(\ref{coincalGF}).
Those lead to an extension of the ``cycle
replacement rule'' to include one-cycles
\cite{rss}:

\begin{equation}
\dot{\cal G}_B(\tau_i,\tau_i)\rightarrow
\dot{\cal G}_B(\tau_i,\tau_i)
-{\cal G}_F(\tau_i,\tau_i)
= {i\over {\rm sin}({\cal Z}){\rm cos}({\cal Z})}
- {i\over {\cal Z}}
\label{onecycle}
\end{equation}

\item

The scalar QED Euler-Heisenberg-Schwinger determinant factor
must be replaced by its spinor QED equivalent,

\be
{\rm det}^{-\half}
\biggl[{\sin({\cal Z})\over {{\cal Z}}}
\biggr] 
\rightarrow
{\rm det}^{-\half}
\biggl[{\tan({\cal Z})\over {{\cal Z} }}
\biggr] 
\label{detchange}
\ee\no

\item

Multiply by the usual factor of $-2$ for statistics and degrees
of freedom.

\end{enumerate}

\subsection{Lorentz covariant decomposition of the 
generalized worldline Green's functions}

For the result to be practically useful
it will be necessary to know
${\cal G}_B,{\cal G}_F$ in more explicit form.
In the calculations performed in \cite{rss,adlsch,frss}
for a purely magnetic field $\bf B$ pointing along the
$z$ - direction the explicit matrix form of 
${\cal G}_B,{\cal G}_F$ had been used. This would
also be possible in the generic case, where, excepting the
case $\bf E\cdot \bf B = 0$, one could use the Lorentz
invariance to choose both $\bf E$ and $\bf B$ to point along
the $z$ - axis. For this case the Green's functions can be
easily written out explicitly. However,
it is possible to directly express them
in terms of Lorentz
invariants, without specialisation of the Lorentz frame.
This can be done in the
following way.
Defining the Maxwell invariants

\bear
f &\equiv& \fourth F_{\mu\nu}F_{\mu\nu}
=
\half (B^2 - E^2)
\non\\
g&\equiv& \fourth F_{\mu\nu}\tilde F_{\mu\nu}
=
i {\bf E}\cdot {\bf B}
\non\\
\label{deffg}
\ear
we have the relations

\bear
F^2 + \tilde F^2
&=&
-2f\Eins
\label{eqF2Ftilde2}\\
F\tilde F
&=&
-g\Eins
\label{eqFFtilde}
\ear

\noindent
Define

\bear
F_{\pm} &\equiv& 
{N_{\pm}^2F-N_+N_-\tilde F\over N_{\pm}^2-N_{\mp}^2}
\label{defFpm}\\
N_{\pm} &\equiv&
\np \pm \nm
\label{defNpm}\\
n_{\pm}
&\equiv&
{\sqrt{f\pm g\over 2}}
\label{defnpm}
\ear
Then one has 

\bear
F &=& F_+ + F_-
\label{decompF}\\
F^2 F_{\pm} &=& -N_{\pm}^2 F_{\pm}
\label{FsquareFpm}\\
F_{+}F_{-}
&=&
0
\label{FpFmorth}
\label{propFpm}
\ear\no
With the help of these relations one easily derives
the following formulas, 

\bear
f_{\rm even}(F) &=&
{1\over N_+^2-N_-^2}
\Bigl\lbrace
-f_{\rm even}
(iN_{+})\bigl[\Nm^2\Eins+F^2\bigr]
+f_{\rm even}(i\Nm)\bigl[\Np^2\Eins + F^2\bigr]\Bigr\rbrace
\non\\
f_{\rm odd}(F) &=&
{i\over N_+^2-N_-^2}
\Bigl\lbrace
\bigl[\Nm f_{\rm odd}(i\Nm) -\Np f_{\rm odd}(i\Np)\bigr] F
\non\\
&&\hspace{50pt}
+
\bigl[\Nm f_{\rm odd}(i\Np)- \Np f_{\rm odd}(i\Nm)\bigr]\tilde F
\Bigr\rbrace
\non\\
\label{formfevenodd}
\ear\no
where $f_{\rm even}$ ($f_{\rm odd}$) are
arbitrary even (odd) functions in the field strength
matrix regular at $F=0$,

\bear
f_{\rm even}(F) = \sum_{n=0}^{\infty}
c_{2n}F^{2n},\qquad
f_{\rm odd}(F) = \sum_{n=0}^{\infty}
c_{2n+1}F^{2n+1}
\label{deffevvenodd}
\ear\no
Decomposing ${\cal G}_{B,F}$ 
into their even (odd) parts
${\cal S}_{B,F}$ (${\cal A}_{B,F}$),

\bear
{\cal G}_{B,F}
&=&
{\cal S}_{B,F}
+
{\cal A}_{B,F}
\label{decomposecalGBGF}
\ear\no
and applying the above formulas we obtain the 
following matrix decompositions
of ${\cal G}_B, \dot {\cal G}_B, {\cal G}_F$,

\bear
{\cal S}_{B12}
&=&
{T\over 2}
\Bigl[
{A^+_{B12}\over z_+}\hat{\cal Z}_+^2
+{A^-_{B12}\over z_-}\hat{\cal Z}_-^2
\Bigr]
\non\\&=&
{T\over 2(z_+^2-z_-^2)}
\Bigl\lbrace
\Bigl[
{z_-^2\over z_+} A_{B12}^{+}
-{z_+^2\over z_-} A_{B12}^{-}
\Bigr]\Eins
+
\Bigl[ {A_{B12}^{+}\over z_+} - {A_{B12}^{-}\over z_-}
\Bigr]{\cal Z}^2
\Bigr\rbrace
\non\\
{\cal A}_{B12}
&=&
{iT\over 2}\Bigl[
(S_{B12}^+ -\dot G_{B12}){\hat{\cal Z_+}\over z_+}
+(S_{B12}^- -\dot G_{B12}){\hat{\cal Z_-}\over z_-}\Bigr]  
\non\\&=&
{iT\over 2(z_+^2-z_-^2)}
\Bigl\lbrace
\bigl[ S_{B12}^{+} - S_{B12}^{-}\bigr] {\cal Z}
+ \Bigl[ {z_+\over z_-} (S_{B12}^{-}-\dot G_{B12}) 
 -{z_-\over z_+} (S_{B12}^{+}-\dot G_{B12})  \Bigr]
{\tilde{\cal Z}}
\Bigr\rbrace
\non\\
\dot{\cal S}_{B12}
&=&
-S^+_{B12}\hat{\cal Z}_+^2
-S^-_{B12}\hat{\cal Z}_-^2
\non\\&=&
{1\over z_+^2-z_-^2}
\Bigl\lbrace
\bigl[
z_+^2 S_{B12}^{-} -z_-^2 S_{B12}^{+} \bigr]\Eins
+
\bigl[ S_{B12}^{-} - S_{B12}^{+}\bigr]{\cal Z}^2
\Bigr\rbrace
\non\\
\dot{\cal A}_{B12}
&=&
-i\Bigl[A_{B12}^-\hat{\cal Z}_-
+ A_{B12}^+\hat{\cal Z}_+\Bigr]
\non\\&=&
{i\over z_+^2-z_-^2}
\Bigl\lbrace
\bigl[ z_{-} A_{B12}^{-} - z_{+} A_{B12}^{+}\bigr] {\cal Z}
+ \bigl[z_- A_{B12}^{+}-z_+ A_{B12}^{-}\bigr] \tilde{\cal Z}
\Bigr\rbrace
\non\\
{\cal S}_{F12} &=&
-S^+_{F12}\hat{\cal Z}_+^2
-S^-_{F12}\hat{\cal Z}_-^2
\non\\&=&
{1\over z_+^2-z_-^2}
\Bigl\lbrace
\bigl[z_+^2 S_{F12}^{-} -z_-^2 S_{F12}^{+}\bigr]
\Eins + \bigl[ S_{F12}^{-} -S_{F12}^{+}\bigr] {\cal Z}^2
\Bigr\rbrace\non\\
{\cal A}_{F12} &=&
-i\Bigl[A_{F12}^-\hat{\cal Z}_- +
A_{F12}^+\hat{\cal Z}_+\Bigr]
\non\\&=&
{i\over z_+^2-z_-^2}
\Bigl\lbrace
\bigl[z_- A_{F12}^{-} -z_+ A_{F12}^{+}\bigr] {\cal Z}
+ \bigl[z_- A_{F12}^{+} -z_+A_{F12}^{-}\bigr] \tilde {\cal Z}
\Bigr\rbrace\non\\ 
\label{decompcalSA}
\ear\no
Here we have further introduced

\bear
z_\pm \equiv eN_\pm T,\quad 
\tilde {\cal Z}\equiv eT\tilde F,\quad
{\cal Z}_{\pm}\equiv
eTF_{\pm}=
{z_{\pm}^2{\cal Z}-z_{+}z_{-}\tilde{\cal Z}\over z_{\pm}^2-z_{\mp}^2}  
,\quad
\hat{\cal Z}_{\pm} \equiv {{\cal Z}_{\pm}\over z_{\pm}}
\non\\
\label{defzs}
\ear\no
Note that ${\cal Z}\tilde{\cal Z}=-z_+z_-\Eins$,
$\hat{\cal Z}_{\pm}^3=-\hat{\cal Z}_{\pm}$.
The scalar, dimensionless 
coefficient functions appearing in these formulas are given by

\bear
S_{B12}^{\pm} &=&
{\sinh(z_{\pm}\dot G_{B12})\over \sinh(z_{\pm})} 
\non\\
A_{B12}^{\pm} &=&
{\cosh(z_{\pm} \dot G_{B12})\over 
\sinh(z_{\pm})}-{1\over z_{\pm}}
\non\\
S_{F12}^{\pm} &=&
G_{F12}{\cosh(z_{\pm}\dot G_{B12})\over\cosh(z_{\pm})}
\non\\
A_{F12}^{\pm} &=&
G_{F12}{\sinh(z_{\pm}\dot G_{B12})\over \cosh(z_{\pm})}
\non\\
\label{defAB}
\ear\no
The non-vanishing coincidence limits are in 
$A_{B,F}^{\pm}$,

\bear
A_{Bii}^{\pm} &=&
\coth(z_{\pm})-{1\over z_{\pm}}
\non\\
A_{Fii}^{\pm} &=&
\tanh(z_{\pm})
\non\\
\label{coinAB}
\ear\no
In the string-inspired formalism, those functions are the basic
building blocks of parameter integrals for processes involving
constant fields. 
Let us also write down the first few terms of the weak field
expansions of these functions,

\bear
S_{B12}^{\pm} &=&
\dot G_{B12}\biggl[
1-{2\over 3}{G_{B12}\over T}z_{\pm}^2
+\Bigl({2\over 45}{G_{B12}\over T}+
{2\over 15}{G_{B12}^2\over T^2}\Bigr)z_{\pm}^4
+\, {\rm O}(z_{\pm}^6)
\biggr]
\non\\
A_{B12}^{\pm} &=&
\Bigl(\third -2{G_{B12}\over T}\Bigr)z_{\pm}
+\Bigl(-{1\over 45}+{2\over 3}
{G_{B12}^2\over T^2}\Bigr)z_{\pm}^3
+\,  {\rm O}(z_{\pm}^5)
\non\\
S_{F12}^{\pm} &=&
G_{F12}\biggl[
1-2{G_{B12}\over T}z_{\pm}^2+{2\over 3}\Bigl({G_{B12}\over T}
+{G_{B12}^2\over T^2}\Bigr)
z_{\pm}^4 +\, {\rm O}(z_{\pm}^6) \biggr]
\non\\
A_{F12}^{\pm} &=&
G_{F12}\dot G_{B12}
\biggl[z_{\pm}-\Bigl(\third + {2\over 3}{G_{B12}\over T}\Bigr)
z_{\pm}^3+\, {\rm O}(z_{\pm}^5)\biggr]
\non\\
\label{expandAB}
\ear\no
(here we used the identity $\dot G_{B12}^2 = 1-{4\over T}G_{B12}$).

\no
In the same way one finds
for the determinant factors 
(\ref{scaldetext}),(\ref{spindetext})

\bear
{\rm det}^{-{1\over 2}}
\biggl[{\sin({\cal Z})\over {{\cal Z}}}
\biggr] &=&
{z_+z_-\over \sinh(z_+)\sinh(z_-)},
\non\\
{\rm det}^{-{1\over 2}}
\biggl[{\tan({\cal Z})\over {{\cal Z}}}
\biggr]
&=&
{z_+z_-\over \tanh(z_+)\tanh(z_-)}
\non\\
\label{decompdet}
\ear\no
In the appendix we write these formulas also for various
special cases of interest.
Those are the magnetic field case ${\bf E}=0$, 
the crossed field case $f=g=0$, and 
the self-dual field $F^{\mu\nu}=\tilde F^{\mu\nu}$.

Using the above formulas we can obtain 
explicit results in a Lorentz covariant way.
Nevertheless, it will be useful
to write down these formulas also for
the Lorentz system where $\bf E$ and $\bf B$ 
are both pointing along the positive z - axis,
${\bf E} = (0,0,E), {\bf B} = (0,0,B)$.
(For this to be possible we have to assume that
${\bf E}\cdot{\bf B} > 0$.) 
In this Lorentz system $g=iEB$, so that

\bear
n_{\pm} = \half(B\pm iE), N_+=B, N_-=iE,
F_+ = Br_{\perp}, F_-= iEr_{\parallel},
\label{special}
\ear\no
where 

\begin{equation}
r_{\perp} \equiv
\left(
\begin{array}{*{4}{c}}
0&1&0&0\\
-1&0&0&0\\
0&0&0&0\\
0&0&0&0
\end{array}
\right),\qquad
r_{\parallel} \equiv
\left(
\begin{array}{*{4}{c}}
0&0&0&0\\
0&0&0&0\\
0&0&0&1\\
0&0&-1&0
\end{array}
\right)\nonumber\\
\label{defr}\nonumber
\non\\
\vspace{5mm}
\end{equation}
\vspace{5 mm}
Introducing also the projectors

\begin{equation}
g_{\perp} \equiv
\left(
\begin{array}{*{4}{c}}
1&0&0&0\\
0&1&0&0\\
0&0&0&0\\
0&0&0&0
\end{array}
\right),\qquad
g_{\parallel} \equiv
\left(
\begin{array}{*{4}{c}}
0&0&0&0\\
0&0&0&0\\
0&0&1&0\\
0&0&0&1
\end{array}
\right)\nonumber\\
\label{defgs}\nonumber
\vspace{7 mm}
\end{equation}
the matrix decompositions (\ref{decompcalSA}) can then
be rewritten as follows,

\bear
{\cal S}_{B12}^{\mu\nu}
&=&
-{T\over 2}
\sum_{\alpha ={\perp},{\parallel}}
{A_{B12}^{\alpha}\over z_{\alpha}}\,g_{\alpha}^{\mu\nu}
\non\\
{\cal A}_{B12}^{\mu\nu}
&=&
{iT\over 2}
\sum_{\alpha ={\perp},{\parallel}}
{S_{B12}^{\alpha}-\dot G_{B12}\over z_{\alpha}}
\,r_{\alpha}^{\mu\nu}
\non\\
\dot{\cal S}_{B12}^{\mu\nu} &=&
\sum_{\alpha ={\perp},{\parallel}}
S_{B12}^{\alpha}\,g_{\alpha}^{\mu\nu}
\non\\
\dot{\cal A}_{B12}^{\mu\nu} &=& 
-i
\sum_{\alpha ={\perp},{\parallel}}
A_{B12}^{\alpha}\,r_{\alpha}^{\mu\nu}
\non\\
{\cal S}_{F12}^{\mu\nu} &=&
\sum_{\alpha ={\perp},{\parallel}}
S_{F12}^{\alpha}\,g_{\alpha}^{\mu\nu}
\non\\
{\cal A}_{F12}^{\mu\nu} &=& 
-i
\sum_{\alpha ={\perp},{\parallel}}
A_{F12}^{\alpha}\,r_{\alpha}^{\mu\nu}
\non\\
\label{specialdecompcalSA}
\ear\no
with
$S/A^{\perp}_{B/F}\equiv S/A^+_{B/F}\,
(z_+=eBT\equiv z_{\perp}),\,
S/A^{\parallel}_{B/F}\equiv S/A^-_{B/F}\,
(z_-=ieET\equiv z_{\parallel})$.

\no
The determinant factors specialize to

\bear
{\rm det}^{-{1\over 2}}
\biggl[{\sin({\cal Z})\over {{\cal Z}}}
\biggr] &=&
{eBTeET\over \sinh(eBT)\sin(eET)}
\non\\
{\rm det}^{-{1\over 2}}
\biggl[{\tan({\cal Z})\over {{\cal Z}}}
\biggr]
&=&
{eBTeET\over \tanh(eBT)\tan(eET)}
\non\\
\label{specialdet}
\ear\no

\section{The Scalar/Spinor
QED Vacuum Polarization Tensors in a Constant Field}
\renewcommand{\theequation}{4.\arabic{equation}}
\setcounter{equation}{0}

We now apply this formalism to
a calculation of the scalar and spinor QED
vacuum polarisation tensors in a general
constant field.
For the 2-point case eq.(\ref{scalarqedmasterF})
yields the following integrand,

\be
\exp\biggl\lbrace
\ldots
\biggr\rbrace
\mid_{\rm multi-linear}\quad
=
\Bigl\lbrack
\varepsilon_1\cdot\ddot {\cal G}_{B12}\cdot\varepsilon_2
-
\varepsilon_1\cdot\dot{\cal G}_{B1i}\cdot k_i
\,\varepsilon_2\cdot\dot{\cal G}_{B2j}\cdot k_j
\Bigr\rbrack
\e^{k_1\cdot \bar{\cal G}_{B12}\cdot k_2}
\label{P2withF}
\ee
where summation over $i,j = 1,2$ is understood.
Removing the second derivative in the first term by a
partial integration in $\tau_1$ 
this becomes

\be
\biggl\lbrack
-\varepsilon_1\cdot\dot {\cal G}_{B12}\cdot\varepsilon_2
\,k_1\cdot \dot {\cal G}_{B1j}\cdot k_j
-
\varepsilon_1\cdot\dot{\cal G}_{B1i}\cdot k_i
\,\varepsilon_2\cdot\dot{\cal G}_{B2j}\cdot k_j
\biggr\rbrack
\e^{k_1\cdot \bar{\cal G}_{B12}\cdot k_2}
\label{P2withFpint}
\ee
We apply the ``cycle replacement rule'' to this expression
and use momentum conservation, $k\equiv k_1 = -k_2$.
The content of the brackets then turns into
$\varepsilon_{1\mu} I^{\mu\nu}\varepsilon_{2\nu}$,
where

\bear
I^{\mu\nu} &=&
\dot{\cal G}^{\mu\nu}_{B12}k\cdot\dot{\cal G}_{B12}\cdot k
-{\cal G}^{\mu\nu}_{F12}k\cdot{\cal G}_{F12}\cdot k
\non\\&&\hspace{-15pt}
- \biggl[
\Bigl(\dot{\cal G}_{B11}
-{\cal G}_{F11}
-\dot{\cal G}_{B12}\Bigr)
^{\mu\lambda}
  \Bigl(\dot{\cal G}_{B21}
-\dot{\cal G}_{B22}
+{\cal G}_{F22}
\Bigr)^{\nu\kappa}
+
{\cal G}^{\mu\lambda}_{F12}
{\cal G}^{\nu\kappa}_{F21}
\biggr]
k^{\kappa}k^{\lambda}
\non\\
\label{substint}
\ear\no
Next we would like to use the fact that this integrand
contains many terms which integrate to zero
due to antisymmetry
under the exchange
$\tau_1\leftrightarrow\tau_2$. 
We therefore decompose ${\cal G}_B$ 
and ${\cal G}_F$ into
their parts symmetric 
and antisymmetric  
in the Lorentz indices
as in (\ref{decomposecalGBGF}).
First note that only the even part
of ${\cal G}_B$ 
contributes in the exponent,

\bear
k_1\cdot \bar {\cal G}_{B12}\cdot k_2
&=&
k_1\cdot \bigl( {\cal S}_{B12}-{\cal S}_{B11}\bigr)\cdot k_2
\equiv
-Tk\cdot\Phi_{12}\cdot k
\non\\
\label{defPhi}
\ear\no
$I^{\mu\nu}$ turns,
after decomposing all factors of $\dot {\cal G}_B, {\cal G}_F$ as above,
and deleting all $\tau$ - odd terms, into 

\bear
I^{\mu\nu}_{\rm spin}
&\equiv&
\biggl\lbrace
\Bigl(
{\dot{\cal S}}^{\mu\nu}_{B12}{\dot{\cal S}}^{\kappa\lambda}_{B12}
- {\dot{\cal S}}^{\mu\lambda}_{B12}{\dot{\cal S}}^{\nu\kappa}_{B12}
\Bigr)
-
\Bigl(
{{\cal S}}^{\mu\nu}_{F12}{{\cal S}}^{\kappa\lambda}_{F12}
- {{\cal S}}^{\mu\lambda}_{F12}{{\cal S}}^{\nu\kappa}_{F12}
\Bigr)
\non\\
&&
+
\Bigl(
{\dot{\cal A}}_{B12}-{\dot{\cal A}}_{B11}+{{\cal A}}_{F11}
\Bigr)^{\mu\lambda}
\Bigl(
{\dot{\cal A}}_{B12}-{\dot{\cal A}}_{B22}+{{\cal A}}_{F22}
\Bigr)^{\nu\kappa}
\non\\
&&
-{\cal A}^{\mu\lambda}_{F12}
{\cal A}^{\nu\kappa}_{F12}
\biggr\rbrace
k^{\kappa}k^{\lambda}
\non\\
\label{intfinal}
\ear\no
(here we used (\ref{symmcalGBF})).
In this way we obtain the following integral representations 
for the dimensionally regularised
scalar/spinor QED vacuum polarisation
tensors \cite{mecorfu}, 

\bear
\Pi^{\mu\nu}_{\rm scal}(k)
&=&
-{e^2\over {[4\pi]}^{D\over 2}}
{\dps\int_{0}^{\infty}}{dT\over T}
{T}^{2-{D\over 2}}
e^{-m^2T}
{\rm det}^{-\half}\biggl[{\sin({\cal Z})\over {{\cal Z}}}
\biggr] 
\int_0^1 du_1
\,\,I^{\mu\nu}_{\rm scal}
\,\e^{-Tk\cdot\Phi_{12}\cdot k}
\non\\
\label{vpscalreg}\\
\Pi^{\mu\nu}_{\rm spin}(k)
&=&
2{e^2\over {[4\pi]}^{D\over 2}}
{\dps\int_{0}^{\infty}}{dT\over T}
{T}^{2-{D\over 2}}
e^{-m^2T}
{\rm det}^{-\half}\biggl[{\tan({\cal Z})\over {{\cal Z}}}\biggr]
\int_0^1 du_1
\,\,I^{\mu\nu}_{\rm spin}
\,\e^{-Tk\cdot\Phi_{12}\cdot k}
\non\\
\label{vpspinreg}
\ear\no
Here $I^{\mu\nu}_{\rm scal}$ is obtained simply by
deleting, in eq. (\ref{intfinal}), all quantities
carrying a subscript ``F''. As usual we have rescaled
to the unit circle and 
set $u_2 =0$. 

Note that again the transversality
of the vacuum polarization tensors is manifest at the
integrand level, 
$k_{\mu}I^{\mu\nu}_{\rm scal/spin}
=
I^{\mu\nu}_{\rm scal/spin}k_{\nu} =0$.

The constant field vacuum polarisation tensors contain
the UV divergences of the ordinary vacuum polarisation tensors
(\ref{scalarvpresult}),(\ref{spinorvpresult}), and thus require
renormalization.
As is usual in this context we perform the
renormalization on-shell, i.e. we impose the following
condition on the renormalized 
vacuum polarization tensor $\bar\Pi^{\mu\nu}(k)$ 
(see e.g. \cite{ditreu}),

\bear
\lim_{k^2\rightarrow 0}\lim_{F\rightarrow 0}
{\bar \Pi}^{\mu\nu}(k) =0
\label{renormcond}
\ear\no
Counterterms appropriate to this
condition are easy to find from our
above results for the ordinary vacuum polarisation
tensors. From the representations
eqs. (\ref{scalvpprel}), (\ref{spinvpprel}) 
for these tensors it is obvious that we can
implement (\ref{renormcond}) by
subtracting those same 
expressions with the last factor $\e^{-G_{B12}k^2}$ 
deleted. In this way we find for the
renormalised vacuum polarisation tensors

\bear
{\bar \Pi}^{\mu\nu}_{\rm scal}(k)
&=&
\Pi^{\mu\nu}_{\rm scal}(k)
+{\alpha\over 4\pi}
\bigl(g^{\mu\nu}k^2 -k^{\mu}k^{\nu}\bigr)
{\dps\int_{0}^{\infty}}{dT\over T}
e^{-m^2T}
\int_0^1 du_1
\dot G_{B12}^2
\non\\
{\bar \Pi}^{\mu\nu}_{\rm spin}(k)
&=&
\Pi^{\mu\nu}_{\rm spin}(k)
-
{\alpha\over 2\pi}
\bigl(g^{\mu\nu}k^2 -k^{\mu}k^{\nu}\bigr)
{\dps\int_{0}^{\infty}}{dT\over T} e^{-m^2T}
\int_0^1 du_1
\bigl(\dot G_{B12}^2-G_{F12}^2\bigr)
\non\\
\label{vpren}
\ear\no
The remaining $u_1$ - integral can be brought
into a more standard form
by a transformation of variables
$v = \dot G_{B12} = 1-2u_1$.
Writing the integrands explicitly
using the formulas (\ref{decompcalSA})
and continuing to Minkoswki space
\footnote{For the Maxwell invariants this means
$f\rightarrow {\cal F}$, 
$g\rightarrow i{\cal G}$,
$N_+\rightarrow a$, $N_-\rightarrow ib$
(to be able to fix all signs we assume 
${\cal G} \geq 0$).
Note also that $r_{\perp}k\rightarrow \tilde k_{\perp},
r_{\parallel}k\rightarrow -i\tilde k_{\parallel}$.
}
we obtain our final result for these amplitudes,

\bear
{\bar \Pi}^{\mu\nu}_{\rm scal}(k)
&=&
-{\alpha\over 4\pi}
{\dps\int_{0}^{\infty}}{ds\over s}
\,e^{-ism^2}
\int_{-1}^1 {dv\over 2}
\Biggl\lbrace
{z_+z_-\over \sinh(z_+)\sinh(z_-)}
\non\\&&\times
{\rm exp}\biggl[
-i{s\over 2}\sum_{\alpha=+,-}
{A_{B12}^{\alpha}-A_{B11}^{\alpha}\over z_{\alpha}}\,
k\cdot \hat{\cal Z}_{\alpha}^2\cdot k
\biggr]
\non\\&&\times
\sum_{\alpha,\beta =+,-}
\biggl(
S_{B12}^{\alpha}
S_{B12}^{\beta}
\Bigl[
\bigl(\hat{\cal Z}_{\alpha}^2\bigr)^{\mu\nu}
k\cdot \hat{\cal Z}_{\beta}^2 \cdot k
-
\bigl(\hat{\cal Z}_{\alpha}^2k\bigr)^{\mu}
\bigl(\hat{\cal Z}_{\beta}^2k\bigr)^{\nu}
\Bigr]
\non\\&&\hspace{55pt}
-
(A_{B12}^{\alpha}-A_{B11}^{\alpha})
(A_{B12}^{\beta}-A_{B22}^{\beta})
\bigl(\hat{\cal Z}_{\alpha}k\bigr)^{\mu}
\bigl(\hat{\cal Z}_{\beta}k\bigr)^{\nu}
\biggr)
\non\\&&
- \bigl(g^{\mu\nu}k^2 -k^{\mu}k^{\nu}\bigr)v^2
\biggr\rbrace
\label{vpscalfinal}\\
{\bar \Pi}^{\mu\nu}_{\rm spin}(k)
&=&
{\alpha\over 2\pi}
{\dps\int_{0}^{\infty}}{ds\over s}
\,e^{-ism^2}
\int_{-1}^1 {dv\over 2}
\Biggl\lbrace
{z_+z_-\over \tanh(z_+)\tanh(z_-)}
\non\\&&\times
{\rm exp}\biggl[
-i{s\over 2}\sum_{\alpha=+,-}
{A_{B12}^{\alpha}-A_{B11}^{\alpha}\over z_{\alpha}}\,
k\cdot \hat{\cal Z}_{\alpha}^2\cdot k
\biggr]
\non\\&&\times
\sum_{\alpha,\beta =+,-}
\biggl(
\Bigl[
S_{B12}^{\alpha}
S_{B12}^{\beta}
-
S_{F12}^{\alpha}
S_{F12}^{\beta}
\Bigr]
\Bigl[
\bigl(\hat{\cal Z}_{\alpha}^2\bigr)^{\mu\nu}
k\cdot \hat{\cal Z}_{\beta}^2 \cdot k
-
\bigl(\hat{\cal Z}_{\alpha}^2k\bigr)^{\mu}
\bigl(\hat{\cal Z}_{\beta}^2k\bigr)^{\nu}
\Bigr]
\non\\&&\hspace{-5pt}
-
\Bigl[ 
(A_{B12}^{\alpha}-A_{B11}^{\alpha}+A_{F11}^{\alpha})
(A_{B12}^{\beta}-A_{B22}^{\beta}+A_{F22}^{\beta})
-A_{F12}^{\alpha}A_{F12}^{\beta}
\Bigr]
\bigl(\hat{\cal Z}_{\alpha}k\bigr)^{\mu}
\bigl(\hat{\cal Z}_{\beta}k\bigr)^{\nu}
\biggr)
\non\\&&
- \bigl(g^{\mu\nu}k^2 -k^{\mu}k^{\nu}\bigr)(v^2-1)
\biggr\rbrace
\label{vpspinfinal}
\ear\no
where now

\bear
z_+ &=& iesa \non\\
z_- &=& -esb \non\\
\hat{\cal Z}_+ &=& {aF-b\tilde F\over a^2+b^2}\non\\
\hat{\cal Z}_- &=& -i{bF+a\tilde F\over a^2 +b^2} \non\\
\label{defsmink}
\ear\no
with $a,b$ denoting 
the standard `secular'
invariants 

\bear
a &\equiv& \sqrt{\sqrt{{\cal F}^2+{\cal G}^2}+{\cal F}}
\non\\
b &\equiv& \sqrt{\sqrt{{\cal F}^2+{\cal G}^2}-{\cal F}}
\label{defsecular}
\ear
(${\cal F} = \half (B^2-E^2), \quad {\cal G} =
{\bf E}\cdot{\bf B}$).

For fermion QED, the 
constant field vacuum polarization tensor was obtained before
by various authors 
\cite{batsha,bakast,urrutia,artimovich,gies}.
For the sake of comparison with their results, let us 
also specialize to the Lorentz system where 
${\bf E}=(0,0,E)$ and ${\bf B}=(0,0,B)$.
In this system $a=B, b=E$.
Denoting

\bear
k_{\parallel} &=& (k^0,0,0,k^3),\quad
k_{\perp} = (0,k^1,k^2,0)\non\\
\tilde k_{\parallel} &=& (k^3,0,0,k^0),\quad
\tilde k_{\perp} = (0,k^2,-k^1,0)
\non\\
\label{defkktilde}
\ear\no
our result can be written as follows,
 
\bear
{\bar \Pi}^{\mu\nu}_{\bigl({{\rm spin}\atop{\rm scal}}\bigr)}(k)
&=&
-{\alpha\over 4\pi}
\biggl({-2\atop 1}\biggr)
{\dps\int_{0}^{\infty}}{ds\over s}
\int_{-1}^1 {dv\over 2}
\Biggl\lbrace
{zz'\over \sin(z)\sinh(z')}
\biggl({\cos(z)\cosh(z')\atop 1}\biggr)
\non\\
&&\times
\e^{-is\Phi_0}
\sum_{\alpha,\beta =\perp ,\parallel}
\Bigl[
s^{\alpha\beta}_{\bigl({{\rm spin}\atop{\rm scal}}\bigr)}
(g^{\mu\nu}_{\alpha}k_{\beta}^2-k^{\mu}_{\alpha}k^{\nu}_{\beta})
+a^{\alpha\beta}_{\bigl({{\rm spin}\atop{\rm scal}}\bigr)}
\tilde k_{\alpha}^{\mu}\tilde k_{\beta}^{\nu}\Bigr]
\non\\
&& -\e^{-ism^2}(g^{\mu\nu}k^2-k^{\mu}k^{\nu})
\biggl({v^2-1\atop v^2}\biggr)
\Biggr\rbrace
\label{specialvpfinal}
\ear\no
where $z=eBs, z'=eEs$, 

\bear
\Phi_0 = m^2 +{k_{\perp}^2\over 2}{\cos(zv)-\cos(z)\over z\sin(z)}
-{k_{\parallel}^2\over 2}{\cosh(z'v)-\cosh(z')\over z'\sinh(z')}
\non\\
\label{Phi0}
\ear

\bear
s_{\rm scal}^{\perp\perp} &=& {\sin^2(zv)\over\sin^2(z)} \non\\
s_{\rm scal}^{\perp\parallel,\parallel\perp}
&=&
{\sin(zv)\sinh(z'v)\over\sin(z)\sinh(z')}\non\\
s_{\rm scal}^{\parallel\parallel} &=&
{\sinh^2(z'v)\over\sinh^2(z')}
\non\\
a_{\rm scal}^{\perp\perp} &=&
\Bigl({\cos(zv)-\cos(z)\over\sin(z)}\Bigr)^2 \non\\
a_{\rm scal}^{\perp\parallel,\parallel\perp}&=&
-{\cos(zv)-\cos(z)\over\sin(z)}{\cosh(z'v)-\cosh(z')\over\sinh(z')}
\non\\
a_{\rm scal}^{\parallel\parallel} &=&
\Bigl({\cosh(z'v)-\cosh(z')\over\sinh(z')}\Bigr)^2 \non\\
\label{coefffinalscal}
\ear

\bear
s_{\rm spin}^{\perp\perp} &=& {\sin^2(zv)\over\sin^2(z)} 
-{\cos^2(zv)\over\cos^2(z)}\non\\
s_{\rm spin}^{\perp\parallel,\parallel\perp}
&=&
{\sin(zv)\sinh(z'v)\over\sin(z)\sinh(z')}
-
{\cos(zv)\cosh(z'v)\over\cos(z)\cosh(z')}
\non\\
s_{\rm spin}^{\parallel\parallel} &=&
{\sinh^2(z'v)\over\sinh^2(z')}
-
{\cosh^2(z'v)\over\cosh^2(z')}
\non\\
a_{\rm spin}^{\perp\perp} &=&
\Bigl({\cos(zv)-\cos(z)\over\sin(z)}-\tan(z)\Bigr)^2 
-{\sin^2(zv)\over\cos^2(z)}\non\\
a_{\rm spin}^{\perp\parallel,\parallel\perp}&=&
-\Bigl({\cos(zv)-\cos(z)\over\sin(z)}-\tan(z)\Bigr)
\Bigl({\cosh(z'v)-\cosh(z')\over\sinh(z')}+\tanh(z')\Bigr)
\non\\&&
-{\sin(zv)\sinh(z'v)\over\cos(z)\cosh(z')}
\non\\
a_{\rm spin}^{\parallel\parallel} &=&
\Bigl({\cosh(z'v)-\cosh(z')\over\sinh(z')}+\tanh(z')\Bigr)^2 
-{\sinh^2(z'v)\over\cosh^2(z')}
\non\\
\label{coefffinalspin}
\ear

\vfill\eject

In this form our integrand for the spinor QED case 
can easily be identified 
with the one given in
\cite{urrutia,gies} using the identities \cite{gies}

\bear
g_{\parallel}^{\mu\nu}k_{\parallel}^2-k_{\parallel}^{\mu}
k_{\parallel}^{\nu}
&=&
-\tilde k_{\parallel}^{\mu}
\tilde k_{\parallel}^{\nu}
\non\\
g_{\perp}^{\mu\nu}k_{\perp}^2-k_{\perp}^{\mu}
k_{\perp}^{\nu}
&=&
\tilde k_{\perp}^{\mu}
\tilde k_{\perp}^{\nu}
\label{idkg}
\ear
and trigonometric identities
\footnote{Up to the global sign, which differs from the one in
\cite{urrutia,gies} due to a different definition of the
vacuum polarization tensor. (Our field theory conventions
follow \cite{weinberg}.)}. 

For the scalar QED case, the vacuum polarization tensor in
a constant field was obtained in \cite{bakast} using the
``operator diagram technique''. The resulting parameter
integral representation agrees with our eq.(\ref{vpscalfinal}).

\section{Discussion}
\renewcommand{\theequation}{5.\arabic{equation}}
\setcounter{equation}{0}

We have described in detail the application of the
`string-inspired' formalism to the calculation of
QED photon amplitudes in a constant external field.
The two-point cases were studied explicitly.
As is usual in the application of the string-inspired
technique to QED processes, our calculation of the
spinor QED vacuum polarisation tensor in a constant field
has yielded the corresponding scalar QED quantity
as a byproduct. 

In both cases we have arrived at compact results written
in terms of the two Lorentz invariants of the
Maxwell field.
Due to our novel
representation (\ref{decompcalSA}) 
of the generalized worldline Green's functions
here this was achieved without the use of explicit
matrix representations.

Already the two-point case clearly shows the advantages
of the method as compared to the standard field theory
techniques used in previous calculations of these
quantities. There are several aspects to this.
The partial integration procedure combined with the
Bern-Kosower `replacement rule' effectively replaces
the calculation of Dirac traces, and is technically
clearly preferable to the latter. Also, the string-inspired
technique combines all loop propagators into a single
propagator from the beginning, while in a Feynman diagram
calculation
each propagator has to be parametrized separately, and
the global loop proper-time variable is to be introduced
at a later stage.
The close relationship between
the scalar and spinor QED calculations 
is also useful. It is due
to the fact that the
``string-inspired'' formalism 
is based on a second-order
field theory formulation of fermion QED 
\cite{berdun,strassler,morgan} rather than the more usual
first-order one.

Clearly one would expect these advantages to become
more significant with increasing number of legs.
For the three-photon amplitude in a magnetic field
this was already demonstrated in \cite{adlsch}.
We believe that in the present formalism even a calculation
of photon-photon scattering in a constant field would
not be excessively cumbersome.

In a sequel paper \cite{ioasch} we will 
consider more generally the calculation of
amplitudes involving both vectors and axial
vectors in a general constant field,
using the generalization of
the path integral representation (\ref{spinorpi})
given in \cite{mcksch,dimcsc}.

Various other generalizations of this formalism
would be of interest. 
An obvious one is the extension to QED amplitudes involving
external scalars or fermions; the relevant generalizations
of eqs.(\ref{scalarqedpi}),(\ref{spinorpi}) have been known
for a long time \cite{fradkin66,casalbuoni,bermar}.
See \cite{dashsu} (scalar QED) and \cite{mckreb,karkto}
(spinor QED) for relevant work for the vacuum case. 
Interesting would also be the
extension to the finite temperature case along the lines
of \cite{shovkovy,haasch,sato}.

\vskip15pt
\noindent{\bf Acknowledgements:}
The author 
would like to thank W. Dittrich, L.E. Hernquist,
M. Reuter and V.I. Ritus
for various helpful informations.

\vfill\eject

\begin{appendix}

\section{Worldline Green's functions for special field
configurations}
\renewcommand{\theequation}{A.\arabic{equation}}
\setcounter{equation}{0}
\vskip10pt

In this appendix we give the explicit form of the
(Euclidean)
generalized worldline Green's functions and
determinants for the following three special
cases of a constant field,
i) the purely magnetic field case,
ii) the crossed - field case, 
iii) the self-dual case.

\begin{enumerate}

\item
{\it Magnetic field case:} \\
$E=g=0,f=\half B^2,n_+=n_-={B\over 2}, N_-=0,N_+=B$

\bear
\bar{\cal S}_{B12} &=& G_{B12}\Eins
+
\biggl[{1\over z^2}G_{B12}
+T{\cosh(z\dot G_{B12})-\cosh(z)\over 2z^3\sinh(z)}
\biggr]{\cal Z}^2
\non\\
{\cal A}_{B12} &=&
{iT\over 2z^2}\Bigl[{\sinh(z\dot G_{B12})\over\sinh(z)}
-\dot G_{B12}\Bigr]{\cal Z}\non\\
\dot{\cal S}_{B12} &=&
\dot G_{B12}\Eins +{1\over z^2}
\Bigl[\dot G_{B12} -{\sinh(z\dot G_{B12})\over\sinh(z)}
\Bigr]{\cal Z}^2
\non\\
\dot {\cal A}_{B12} &=&
i\Bigl[{1\over z^2}-{\cosh(z\dot G_{B12})\over z\sinh(z)}
\Bigr]\,{\cal Z}\non\\
{\cal S}_{F12} &=& G_{F12}\Eins +
G_{F12}\Bigl[{1\over z^2}-{\cosh(z\dot G_{B12})\over z^2\cosh(z)}
\Bigr] {\cal Z}^2\non\\
{\cal A}_{F12} &=&
-iG_{F12}{\sinh(z\dot G_{B12})\over z\cosh(z)}\,{\cal Z}\non\\
\label{decompcalSAmag}
\ear\no 

\bear
{\rm det}^{-{1\over 2}}
\biggl[{\sin({\cal Z})\over {\cal Z}}
\biggr]
&=&
{z\over\sinh(z)},
\qquad
{\rm det}^{-{1\over 2}}
\biggl[{\tan({\cal Z})\over {\cal Z}}
\biggr] =
{z\over\tanh(z)}
\non\\
\label{detmag}
\ear\no
where $z=eBT$.
We have subtracted 
from ${\cal S}_B$ its coincidence limit,
which is expressed by the ``bar''.
The purely electric case is, of course, analogous.

\item
{\it Crossed field case:}\\
$f=g=n_{\pm}=N_{\pm}=0$

\bear
\bar{\cal S}_{B12} &=&
G_{B12}\Eins + {1\over 3T}G_{B12}^2{\cal Z}^2\non\\
{\cal A}_{B12} &=&
-{i\over 3}\dot G_{B12}G_{B12}{\cal Z}\non\\
\dot{\cal S}_{B12} &=&
\dot G_{B12}\Eins +{2\over 3T}\dot G_{B12}G_{B12}{\cal Z}^2\non\\
\dot{\cal A}_{B12} &=&
i\Bigl[{2\over T}G_{B12}-\third\Bigr]{\cal Z}\non\\
{\cal S}_{F12} &=&
G_{F12}\Eins + {2\over T}G_{F12}G_{B12}{\cal Z}^2\non\\
{\cal A}_{F12} &=&
-iG_{F12}\dot G_{B12}{\cal Z}\non\\
\label{decompcalSAcrossed}
\ear\no

\bear
{\rm det}^{-{1\over 2}}
\biggl[{\sin({\cal Z})\over {\cal Z}}
\biggr]
&=&
{\rm det}^{-{1\over 2}}
\biggl[{\tan({\cal Z})\over {\cal Z}}
\biggr] = 1
\label{detcrossed}
\ear\no

\item
{\it Self dual case:}\\
$F=\tilde F,f=g,n_-=0,n_+=N_+=N_-=\sqrt{f}$

\bear
\bar{\cal S}_{B12} &=&
{T\over 2}{\cosh(Z)-\cosh(Z\dot G_{B12})\over Z\sinh(Z)}
\Eins\non\\
{\cal A}_{B12} &=&
i{T\over 2Z^2}\Bigl[{\sinh(Z\dot G_{B12})\over\sinh(Z)}
-\dot G_{B12}\Bigr]{\cal Z}\non\\
\dot{\cal S}_{B12} &=&
{\sinh(Z\dot G_{B12})\over\sinh(Z)}\Eins\non\\
\dot{\cal A}_{B12} &=&
-i\Bigl[{\cosh(Z\dot G_{B12})\over Z\sinh(Z)}-{1\over Z^2}\Bigr]
{\cal Z}
\non\\
{\cal S}_{F12} &=&
G_{F12}{\cosh(Z\dot G_{B12})\over\cosh(Z)}\Eins
\non\\
{\cal A}_{F12} &=&
-iG_{F12}{\sinh(Z\dot G_{B12})\over Z\cosh(Z)}{\cal Z}
\non\\
\label{decompcalSAselfdual}
\ear\no

\bear
{\rm det}^{-{1\over 2}}
\biggl[{\sin({\cal Z})\over {\cal Z}}
\biggr]
&=&
{Z^2\over\sinh^2(Z)},
\qquad
{\rm det}^{-{1\over 2}}
\biggl[{\tan({\cal Z})\over {\cal Z}}
\biggr] =
{Z^2\over\tanh^2(Z)}
\non\\
\label{detselfdual}
\ear\no
where $Z = \sqrt{f}eT$.

\end{enumerate}

\end{appendix}

\vfill\eject


\begin{thebibliography}{99}
\bibitem{eulkoc}
H. Euler and B. Kockel, Naturwissenschaften {\bf 23}
(1935) 246.
\bibitem{eulhei}
W. Heisenberg and H. Euler, Z. Phys. {\bf 98} (1936) 714.
\bibitem{weisskopf}
V. Weisskopf, K. Dan. Vidensk. Selsk. Mat. Fy. Medd. {\bf 14}
(1936) 1,
 reprinted in {\it Quantum Electrodynamics},
J. Schwinger (Ed.) (Dover, New York, 1958).
\bibitem{ritusbook}
V. I. Ritus, {\sl The Lagrangian Function of
an Intense Electromagnetic Field and Quantum Electrodynamics
at Short Distances}, in {\sl Proc. Lebedev Phys. Inst. vol. 168},
V. I. Ginzburg ed., Nova Science Publ., NY 1987.
\bibitem{ditreu}
W. Dittrich and M. Reuter, {\it Effective Lagrangians in
Quantum Electrodynamics}, (Springer, 1985).
\bibitem{grereibook}
W. Greiner and J. Reinhardt, {\it Quantum Electrodynamics}
(Springer, 1992).
\bibitem{ditgie}
W. Dittrich and H. Gies, ``Vacuum Birefringence in Strong Magnetic Fields'',
Proc. of {\sl Frontier Tests of
Quantum Electrodynamics and Physics of the Vacuum}, Sandansky 1998,
1998 Heron Press, 29, (hep-ph/9806417).
\bibitem{pvlas}
R. Pengo et al., Workshop Sandansky 1998, loc. cit., 59.
\bibitem{fock}
V. Fock, Physik. Z. Sowjetunion {\bf 12} (1937) 404.
\bibitem{schwinger51}
J. Schwinger, Phys. Rev. {\bf 82} (1951) 664.
\bibitem{toll}
J. Toll, PhD thesis, Princeton University 1952 (unpublished).
\bibitem{minguzzi}
A. Minguzzi, Nuovo Cim. {\bf 6} (1957) 501;
{\bf 9} (1958) 145; {\bf 19} (1961) 847.
\bibitem{baibre}
R. Baier and P. Breitenlohner, Acta Phys. Austr. {\bf 25} (1967) 
212; Nuovo Cim. {\bf 47} (1967) 261.
\bibitem{constantinescu}
D.H. Constantinescu, Nucl. Phys. {\bf B 36} (1972) 121.
\bibitem{tsaerb}
W.-Y. Tsai and T. Erber, Phys. Rev. {\bf D 10} (1974) 492.
\bibitem{covkal}
R.A. Cover and G. Kalman, Phys. Rev. Lett. {\bf 33} (1974) 1113.
\bibitem{shabad}
A.E. Shabad, Ann. Phys. {\bf 90} (1975) 166.
\bibitem{narozhnyi}
N.B. Narozhnyi, Sov. Phys. JETP {\bf 28} (1969) 371.
\bibitem{ritus72}
V.I. Ritus, Ann. Phys. {\bf 69} (1972) 555.
\bibitem{batsha} 
I.A. Batalin and A.E. Shabad, Sov. Phys. JETP {\bf 33} (1971) 483.
\bibitem{bakast}
V. N. Baier, V. M. Katkov, and V. M. Strakhovenko,
Sov. Phys. JETP 40 (1975) 225; 
Sov. Phys. JETP 41 (1975) 198.
\bibitem{urrutia}
L.F. Urrutia, Phys. Rev. {\bf D 17} (1978) 1977.
\bibitem{artimovich}
G.K. Artimovich, Sov. Phys. JETP {\bf 70} (1990) 787.
\bibitem{gies}
H. Gies, PhD thesis, T{\"u}bingen University (1999);
hep-ph/9909500.
\bibitem{biabia} 
Z. Bialynicka-Birula and I. Bialynicka-Birula,
Phys. Rev. {\bf D 2}
(1970) 2341.
\bibitem{adler71} 
S.L. Adler, Ann. Phys. NY {\bf 67} (1971) 599.
\bibitem{raffelt}
G.G. Raffelt, {\em Stars as Laboratories for Fundamental Physics},
University of Chicago Press, Chicago 1996.
\bibitem{barhar}
M. G. Baring and A. K. Harding 
(astro-ph/9704210).
\bibitem{kaspi}
V.M. Kaspi et al.,
to appear in the Proc. of
{\it Stellar Endpoints, AGN and the Diffuse Background},
Bologna 1999.
\bibitem{bakalov}
D. Bakalov, INFN/AE-94/27 (unpublished).
\bibitem{fermilab877}
S.A. Lee et al., ``Measurement of the Magnetically-induced QED
Birefringence of the Vacuum and an Improved Laboratory Search
for Light Pseudoscalars'', FERMILAB-PROPOSAL-P-877A, 1998.
\bibitem{geheniau}
J. G{\'e}h{\'e}niau, Physica {\bf 16} (1950) 822.
\bibitem{tsai}
W. Y. Tsai, Phys. Rev. {\bf D 10} (1974) 1342.
\bibitem{berkos} 
Z. Bern, D.A. Kosower, Phys. Rev. {\bf D 38} (1988) 1888;
Phys. Rev. Lett. {\bf 66} (1991) 1669;
Nucl. Phys. {\bf B379} (1992) 451.
\bibitem{5glu}
Z. Bern, L. Dixon, D. A. Kosower,
  Phys. Rev. Lett. {\bf 70} (1993) 2677 (hep-ph/9302280).
\bibitem{bedush}
Z. Bern, D. C. Dunbar, and T. Shimada, Phys. Lett. {\bf
  B312}(1993) 277 (hep-th/9307001).
\bibitem{dunnor} 
D.C. Dunbar and P. S. Norridge, 
Nucl. Phys. {\bf B 433} (1995) 181 (hep-th/9408014).
\bibitem{berdun}
Z. Bern and D. C. Dunbar, Nucl. Phys. {\bf B379} (1992)
562.
\bibitem{strassler}
M. J. Strassler, Nucl. Phys. {\bf B385} (1992) 145.
\bibitem{ss3}
M.G. Schmidt and C. Schubert, Phys. Rev. {\bf D53}
(1996) 2150 (hep-th/9410100).
\bibitem{ss1}
M. G. Schmidt and C. Schubert, Phys. Lett. {\bf B318}
  (1993) 438 (hep-th/9309055).
\bibitem{cadhdu}
D. Cangemi, E. D'Hoker, and G. Dunne, 
Phys. Rev. {\bf D 51} (1995) 2513.
\bibitem{gussho}
V.P. Gusynin and I.A. Shovkovy,
J. Phys. {\bf 74} (1996) 282 (hep-ph/9509383);
UCTP-106-98 (hep-th/9804143). 
\bibitem{shaisultanov}
R. Shaisultanov, Phys. Lett. {\bf B 378} (1996) 354 (hep-th/9512142).
\bibitem{rss}
M. Reuter, M.G. Schmidt, and C. Schubert,
Ann. Phys. {\bf 259}, 313 (1997) (hep-th/9610191). 
\bibitem{adlsch}
S. L. Adler and C. Schubert, Phys. Rev. Lett.
{\bf 77}, 1695 (1996) (hep-th/9605035). 
\bibitem{shovkovy}
I.A. Shovkovy, Phys. Lett. {\bf B441} (1998) 313 (hep-th/9806156). 
\bibitem{fhss}
D. Fliegner, M.G. Schmidt, and C. Schubert,
Z. Phys. {\bf C 64} (1994) 111 (hep-ph/9401221);
and with P. Haberl, Ann. Phys. (N.Y.) {\bf 264} (1998) 
51 (hep-th/9707189).
\bibitem{ss2}
M. G. Schmidt and C. Schubert, Phys. 
Lett. {\bf B331} (1994) 69 (hep-th/9403158).
\bibitem{dashsu}
K. Daikouji, M. Shino, and Y. Sumino, 
Phys. Rev. {\bf D53} (1996) 4598
(hep-ph/9508377). 
\bibitem{rolsat}
K. Roland and H.-T. Sato, 
Nucl. Phys. {\bf B480} 99 (1996) (hep-th/9604152).
\bibitem{frss}
D. Fliegner, M. Reuter, M.G. Schmidt, and C. Schubert,
Theor. Math. Phys. {\bf 113} (1997) 1442 (hep-th/9704194).
\bibitem{korsch}
B. K{\"o}rs and M.G. Schmidt, Eur. Phys. J. {\bf C6} (1999) 175 
(hep-th/9803144).
\bibitem{dunsch}
G. V. Dunne and C. Schubert, Nucl. Phys.
{\bf B 564} (2000) 591 (hep-th/9907190). 
\bibitem{berntasi}
Z. Bern, TASI Lectures,
Boulder TASI 92, 471 (hep-ph/9304249).
\bibitem{zako}
C. Schubert, Lectures given at
36th Cracow School of Theoretical Physics, Zakopane 1996,
Act. Phys. Polon. {\bf B 27} (1996) 3965 (hep-th/9610108).
\bibitem{fried}
H.M. Fried, {\it Functional Methods and Models in Quantum Field 
Theory}, MIT Press, Cambridge (1972).
\bibitem{baboca}
A. Barducci, F. Bordi, and R. Casalbuoni,
Nuovo Cim. {\bf 64 B} (1981) 287.
\bibitem{bafrsh}
I.A. Batalin, E.S. Fradkin, and Sh.M. Shvartsman,
Nucl. Phys. {\bf B258} (1985) 435. 
\bibitem{rajeev}
S.G. Rajeev, Ann. Phys. {\bf 173} (1987) 249.
\bibitem{frgish}
E.S. Fradkin, D.M. Gitman, and 
S.M. Shvartsman, {\sl Quantum Electrodynamics
with Unstable Vacuum}, Springer 1991.
\bibitem{mckshe}
D.G.C. McKeon and T.N. Sherry,
Mod. Phys. Lett. {\bf A9} (1994) 2167.
\bibitem{feyn}
R. P. Feynman, Phys. Rev. {\bf 80} (1950) 440.
\bibitem{grscwi}
M.B. Green, J.H. Schwarz, and E. Witten, 
{\it Superstring Theory}, Cambridge University Press 1987.
\bibitem{itzzub}
C. Itzykson and J. Zuber, {\it Quantum Field Theory},
McGraw-Hill 1985.
\bibitem{fradkin66}
E. S. Fradkin, Nucl. Phys. {\bf 76} (1966) 588.
\bibitem{casalbuoni}
R. Casalbuoni, Nuov. Cim. {\bf 33A} (1976) 389.
\bibitem{bermar}
F.A. Berezin and M.S. Marinov, Ann. Phys. {\bf 104} (1977) 336.
\bibitem{mepartint}
C. Schubert, Eur. Phys. J. {\bf C5} (1998) 693
(hep-th/9710067).
\bibitem{mecorfu}
C. Schubert, LAPTH-Conf-729/99, to appear in
the Proc. of {\sl Corfu Summer Institute on Elementary
Particle Physics, 1998} (hep-ph/9905525).
\bibitem{weinberg}
S. Weinberg, {\sl The Quantum Theory of Fields},
Vol. 1, Cambridge Univ. Pr. 1995.
\bibitem{morgan}
A. Morgan, Phys. Lett. {\bf B351} (1995) 249
(hep-ph/9502230).
\bibitem{ioasch}
C. Schubert, in preparation.
\bibitem{mcksch}
D.G.C. McKeon and C. Schubert, Phys. Lett. 
{\bf B440} (1998) 101 (hep-th/9807072).
\bibitem{dimcsc}
F.A. Dilkes, D.G.C. McKeon, and C. Schubert,
J. High Energy Phys. {\bf 03} (1999) 022 (hep-th/9812213). 
\bibitem{mckreb}
D.G.C. McKeon and A. Rebhan,
Phys. Rev. {\bf D48} (1993) 2891.
\bibitem{karkto}
A.I. Karanikas and C.N. Ktorides, JHEP 9911:033 (hep-th/9905027). 
\bibitem{haasch}
M. Haack and M.G. Schmidt, 
Eur. Phys. J. {\bf C7} (1999) 149 (hep-th/9806138). 
\bibitem{sato}
H.-T. Sato, HD-THEP-98-41 (hep-th/9809053).
\end{thebibliography}
\end{document}